\documentclass[twocolumn,english,prl,showpacs]{revtex4-1}
\usepackage[colorlinks=true,urlcolor=blue,citecolor=blue,linkcolor=blue,breaklinks=true]{hyperref}
\usepackage[T1]{fontenc}
\usepackage[latin9]{inputenc}
\usepackage{amssymb}
\usepackage{graphicx}
\usepackage{amsmath,color}
\usepackage{mathrsfs}
\usepackage{amsthm}
\usepackage{float}
\usepackage{braket}
\usepackage{indentfirst}
\usepackage{cases}

\newtheoremstyle{dotless}{}{}{\itshape}{}{\bfseries}{}{ }{}
\theoremstyle{dotless}
\newtheorem*{theorem*}{Theorem}
\newtheorem*{corollary*}{Corollary}
\newtheorem*{conjecture*}{Conjecture}

\newcommand{\Eq}[1]{Eq.~(\ref{#1})}
\newcommand{\Tr}{{\rm Tr}}
\newcommand{\diag}{{\rm diag}}
\newcommand{\acosh}{{\rm acosh}}
\newcommand{\acos}{{\rm acos}}

\begin{document}

\title{Split orthogonal group:\\ A guiding principle for sign-problem-free fermionic simulations}

\author{Lei Wang$^{1}$, Ye-Hua Liu$^{1}$, Mauro Iazzi$^{1}$, Matthias Troyer$^{1}$ and Gergely Harcos$^{2}$}
\affiliation{$^{1}$Theoretische Physik, ETH Zurich, 8093 Zurich, Switzerland}
\affiliation{$^{2}$Alfr\'ed R\'enyi Institute of Mathematics, Re\'altanoda utca 13-15., Budapest H-1053, Hungary}
\begin{abstract}
We present a guiding principle for designing fermionic Hamiltonians and quantum Monte Carlo (QMC) methods that are free from the infamous sign problem by  exploiting the Lie groups and Lie algebras that appear naturally in the Monte Carlo weight of fermionic QMC simulations. Specifically, rigorous mathematical constraints on the determinants involving matrices that lie in the \emph{split orthogonal group} provide a guideline for  sign-free simulations of fermionic models on bipartite lattices. This guiding principle not only unifies the recent solutions of the sign problem based on the continuous-time quantum Monte Carlo methods and the Majorana representation, but also suggests new efficient algorithms to simulate physical systems that were previously prohibitive because of the sign problem. 
\end{abstract}


\maketitle



One of the biggest challenges to classical simulation of quantum systems is the infamous fermion sign problem of quantum Monte Carlo (QMC) simulations. It appears when the weights  of configurations in a QMC simulation may become  negative and therefore cannot be directly interpreted as probabilities~\cite{LohJr:1990up}.  In the presence of a  sign problem, the simulation effort typically grows exponentially with system size and inverse temperature.

While the sign problem is nondeterministic polynomial (NP) hard~\cite{Troyer:2005hv}, implying that there is little hope of finding a \emph{generic} solution, this does not exclude \emph{ad hoc} solutions to the sign problem for specific  models. For example,  one can sometimes exploit symmetries to design appropriate sign-problem-free QMC algorithms for a restricted class of models~\cite{PhysRevLett.83.3116}. However, it is unclear how broad these classes are and it is in general hard to foresee whether a given physical model would have a sign problem in \emph{any} QMC simulations. The situation is not dissimilar to the study of many intriguing problems in the NP complexity class, where a seemingly infeasible problem might turn out to have a polynomial-time solution surprisingly~\cite{Hayes:2008cm}. 

A fruitful approach in pursuing such specific solutions is to design Hamiltonians that capture the right low energy physics and allow sign-problem-free QMC simulations at the same time, called ``designer'' Hamiltonians~\cite{Kaul:2013ika}. This naturally calls for design principles. For bosonic and quantum spin systems a valuable guiding principle is the Marshall sign rule~\cite{Marshall:1955tw, Kaul:2015hg} which  ensures nonnegative weight for all configurations. The design of the sign-problem-free fermionic Hamiltonians  is harder. The method of choice for fermionic QMC simulations are the determinantal QMC approaches,  including traditional discrete-time~\cite{Blankenbecler:1981vj} and  new continuous-time approaches~\cite{1999PhRvL..82.4155R, Rubtsov:2005iw, Gull:2008cm, Iazzi:2014vv, Wang:2015tf}. Both approaches map the original interacting system to free fermions with an imaginary-time dependent Hamiltonian. The partition function is then written as a weighted sum of matrix determinants after tracing out the fermions~\cite{Blankenbecler:1981vj,1999PhRvL..82.4155R, Iazzi:2014vv}:
\begin{equation}
Z = \sum_{\mathcal{C}}f_{\mathcal{C}} \, \det\left[ I + \mathcal{T}e^{-\int_{0}^{\beta}d\tau H_{\mathcal{C}}(\tau)} \right], 
\label{eq:general}
\end{equation}
where $f_{\mathcal{C}}$ is a c-number and $H_{\mathcal{C}}(\tau)$ is an imaginary-time dependent single-particle Hamiltonian matrix (whose matrix elements denote hopping amplitudes and onsite energies on a lattice), both depending on the Monte Carlo configuration $\mathcal{C}$. $\mathcal{T}$ denotes the time-ordering and $I$ is the identity matrix.  
The appearance of the matrix determinant complicates the analysis of the sign problem because it is often not straightforward to see the sign of the Monte Carlo weight of a given configuration~\cite{Batrouni:1993tn, Gubernatis:1994ta}, and the sign of the determinant is related  \cite{Iazzi:2014uo} to the Aharonov-Anandan phase~\cite{PhysRevLett.58.1593} of the imaginary-time evolution. 
The situation is further complicated by the fact that  even for a given physical model the choice of the effective Hamiltonian $H_{\mathcal{C}}$ is not unique (it depends on details of the QMC algorithm such as whether and how to perform an auxiliary field decomposition) and the specific choice may affect the appearance of the sign problem~\cite{Batrouni:1990ug,Batrouni:1993tn,Chen:1992wo}. 

One successful guiding principle for fermionic simulations that has been discovered in the context of nuclear physics~\cite{PhysRevC.48.1518, Anonymous:JutiR0Zk}, lattice QCD~\cite{Hands:2000kq} and condensed matter physics~\cite{Wu:2005im} relies on the time-reversal-symmetry (TRS) of the effective Hamiltonian $H_{\mathcal{C}}$. TRS ensures a nonnegative matrix determinant in \Eq{eq:general} because the eigenvalues of the matrix necessarily appear in Kramers pairs. 
A typical example of this kind is the attractive Hubbard model at balanced filling of two spin species, where after decomposition of the interaction term the Monte Carlo weight even factorizes into product of two identical matrix determinants. Additional conditions such as half filling and bipartiteness of the lattice lead to a solution of the sign problem for the repulsive Hubbard model.
See Refs.~\cite{Wu:2005im, Assaad:2008hx} for a thorough discussion and Refs.~\cite{Hohenadler:2011kk, Berg:2012ie, Wang:2014gx} for several recent applications of the TRS principle.  

Unfortunately, besides the quite intuitive TRS principle~\cite{Anonymous:JutiR0Zk, Hands:2000kq, Wu:2005im}, a broad  criterion for the sign of the matrix determinant is still lacking. Recent progresses on solving the sign problem in a class of fermionic models using the continuous-time quantum Monte Carlo approach~\cite{Huffman:2014fj} and the Majornara representation~\cite{Li:2015jf} provide hints about such a guiding principle. For example, one could search for real-antisymmetric matrices with nonnegative determinant~\cite{Huffman:2014fj, Wang:2014iba}, or try to split the fermionic operator into Majorana fermions for a potential cancellation of the sign~\cite{Li:2015jf}. However, compared to the TRS principle~\cite{Anonymous:JutiR0Zk, Hands:2000kq, Wu:2005im}, both approaches are still not enlightening enough to serve as a guiding principle. Moreover, because of the different appearances of the two solutions~\cite{Huffman:2014fj, Li:2015jf}, it is unclear what is the connection between them and whether there is a deeper underlying reason for such solutions. 

In this Letter, we present a guiding principle that not only unifies the two recent solutions to the sign problem~\cite{Huffman:2014fj,Li:2015jf}, but also suggests a general strategy that enables us to discover solutions to the sign problem for a broader class of fermionic models. The guiding principle exploits the symmetry of the \emph{effective Hamiltonian} $H_{\mathcal{C}}$ and consequently the Lie group structure of the \emph{evolution matrix} $\mathcal{T}e^{-\int_{0}^{\beta} d\tau H_{\mathcal{C}}}$. In particular, the \emph{split orthogonal group} $O(n,n)$ is formed by 
all $2n\times 2n$ real matrices that preserve the metric $\eta = \diag(\underbrace{1,\ldots,1}_{n}, \underbrace{-1,\ldots,-1}_{n})$ 
\begin{equation}
M^{T}\eta M = \eta. 
\label{eq:group}
\end{equation}
Similar to the Lorentz group $O(3,1)$, a familiar example in  relativistic physics, the $O(n,n)$ group contains \emph{four} components. More explicitly, writing the matrix $M$ in the form $M=\left(\begin{array}{cc}M_{11} & M_{12} \\M_{21} & M_{22}\end{array}\right) $ with $n\times n$ sub-blocks, one has $|\det(M_{11})|\ge 1 $ and $|\det(M_{22})|\ge 1 $~\cite{nishikawa1983exponential}.  The four components of $O(n,n)$ can be classified by the signs of $\det(M_{11})$ and $\det(M_{22})$, denoted as $O^{\pm\pm}(n,n)$. Different components can only be connected by \emph{improper} rotations that change the sign of the determinant of the sub-block $M_{11}$ or $M_{22}$. Only the $O^{++}(n,n)$ component forms a subgroup because it contains the identity element.  
\begin{theorem*}
If $M$ belongs to the split orthogonal group $O(n,n)$, then the following statements hold~\cite{SM,MO,Taoblog} 
  \begin{subequations}
    \begin{numcases}
      {\det\left(I+M\right)}
     \ge 0, & \text{if $M\in O^{++}(n,n)$}, \label{eq:theorem++}\\
     \le 0, & \text{if $M\in O^{--}(n,n)$}, \label{eq:theorem--}\\
      = 0, & \text{otherwise}. \label{eq:theorem+-}  
    \end{numcases}
    \label{eq:theorem}
  \end{subequations}
\end{theorem*}
This rather strong statement about the definite sign of the matrix determinant, no matter positive or negative, is invaluable for the determinantal QMC simulations. Furthermore, we have the following  

\begin{corollary*}
Given an arbitrary number of real matrices 
$A_{i}$ that satisfy $\eta A_{i}\eta =-A_{i}^{T}$, we have~\cite{SM,MO} 
\begin{equation}\det\left(I + \prod_{i}e^{A_{i}}\right)\ge 0. 
\label{eq:corollary}
\end{equation}
\end{corollary*}
The proof follows immediately by noticing that $A_{i}$ lies in the Lie algebra of the group $O(n,n)$~\cite{orthogonalgroup}. Each factor of the matrix product $\prod_{i}e^{A_{i}}$ is an exponential from the Lie algebra to the Lie group $O(n,n)$ in the identity component, thus \Eq{eq:corollary} is a consequence of \Eq{eq:theorem++}. Note that the form of the matrix determinant of \Eq{eq:corollary} resembles the weight that appears in the determinantal QMC calculations \Eq{eq:general}~\cite{Blankenbecler:1981vj,1999PhRvL..82.4155R, Iazzi:2014vv}. 

Before moving on, we comment on the general relevance of \Eq{eq:theorem} and \Eq{eq:corollary} to physical problems. On a bipartite lattice, the parities of the sublattices naturally provide the metric $\eta$ appearing in \Eq{eq:group}. To further reveal its physical meaning, we write an element in the Lie algebra $A_{i} = \left(\begin{array}{cc} C_{i} & B_{i} \\B^{T}_{i} & D_{i}\end{array}\right)$ explicitly. In the special case of $C_{i}=D_{i}=0$, $A_{i}$ can be recognized as a bipartite single-particle Hamiltonian and the condition on $A_{i}$ has appeared in Eq.~(4) of Ref.~\cite{Huffman:2014fj}. The Corollary (\ref{eq:corollary}) states that the partition function of such bipartite imaginary-time dependent noninteracting system is nonnegative~\footnote{This fact was used implicitly in the CT-QMC calculation of the R\'enyi entanglement entropy in~\cite{Wang:2014ir}.}. Moreover, in general the matrix $A_{i}$ does not need to be symmetric. The condition on $A_{i}$ only requires $C_{i}^{T}= -C_{i}$ and $D_{i}^{T} = -D_{i}$, thus provides more flexibilities in designing the QMC approaches. 

To see how the above rigorous mathematical statements apply to  determinantal QMC simulations of physical systems, we  consider first the spinless $t-V$ model on a bipartite lattice,
\begin{equation}
 \hat{H}= { \sum_{i,j} \hat{c}^{\dagger}_{{i}} K_{ij} \hat{c}_{{j}}} + \sum_{\braket{i,j}} { \left[V\left( \hat{n}_{i}\hat{n}_{j} - \frac{\hat{n}_{i}+\hat{n}_{j}}{2}\right)  -\Gamma \right]  }. 
\label{eq:tVmodel}
\end{equation}
Here $\hat{c}^\dagger_{i}$ and $\hat{c}_{i}$ are fermion creation and annihilation operators and $\hat{n}_{i}=\hat{c}^\dagger_{i}\hat{c}_{i}$ is the occupation number operator on site $i$. There are $2n$ lattice sites which split into two sublattices $\mathcal{A}$ and $\mathcal{B}$. In accordance with the metric $\eta$,  we sort the sites by placing all sites in $\mathcal{A}$ before those in $\mathcal{B}$. The bipartite hopping matrix $K$ has zeros on the diagonal and is real-symmetric, therefore fulfills the requirement $\eta K \eta = -K^{T}$ of the Corollary~\cite{Huffman:2014fj}. The second term is a repulsive interaction between nearest-neighbors $\braket{i,j}$ (belonging to different sublattices) and we introduced a constant shift $\Gamma$, which will play a crucial role in later discussions. 

We employ the continuous-time quantum Monte Carlo (CT-QMC) framework~\cite{1999PhRvL..82.4155R, Rubtsov:2005iw, Gull:2008cm, Iazzi:2014vv} in the following analysis. This approach is free from time discretization errors, as efficient~\cite{Iazzi:2014vv} and more flexible and powerful~\cite{Wang:2015tf, Wang:2015twa} than the discrete-time counterpart~\cite{Blankenbecler:1981vj}. Furthermore the discrete-time algorithms can be derived as a restricted version of CT-QMC on an equidistant grid of imaginary-times~\cite{Mikelsons:2009eka} and our results apply to them as well. We rewrite the Hamiltonian \Eq{eq:tVmodel} as $\hat{H} = \hat{H}_{0}+\sum_{\braket{i,j} }\hat{v}_{ij}$ and perform an expansion in the interaction term~\cite{1999PhRvL..82.4155R}
\begin{widetext}
\begin{eqnarray}
Z = \Tr \left(e^{-\beta \hat{H}}\right) 
= \sum_{k=0}^{\infty} \sum_{\braket{i_{1},j_{1}}} \ldots \sum_{\braket{i_{k},j_{k}}} \int_{0}^{\beta}d \tau_{1} \ldots \int^{\beta}_{\tau_{k-1}} d\tau_{k}\, \Tr\left[e^{-(\beta-\tau_{k})\hat{H}_{0} } \left(-\hat{v}_{i_{k}j_{k}}\right) \ldots \left(-\hat{v}_{i_{1}j_{1}}\right) e^{-\tau_{1}\hat{H}_{0}} \right].  \label{eq:expansion} 
\end{eqnarray}
\end{widetext}
At this point there are multiple ways to proceed, which result in distinct CT-QMC algorithms, differing in efficiency and the use of auxiliary fields (see \cite{Wang:2015tf} for an overview). In particular, we choose the following auxiliary field decomposition of the interaction term to reveal the connections of various solutions to the sign problem~\cite{Huffman:2014fj, Li:2015jf},   
\begin{equation}
 - \hat{v}_{ij} = \frac{\Gamma}{2} \sum_{\sigma =\pm} \exp \left[\sigma \lambda \left(\hat{c}_i^{\dagger} \hat{c}_j+ \hat{c}_j^{\dagger} \hat{c}_i \right)\right], 
 \label{eq:auxfield+}
\end{equation}
where $\lambda = \acosh \left(1 + \frac{V}{2 \Gamma}\right) $ is a real number for \emph{repulsive} interaction $V$ and \emph{positive} shift $\Gamma$. The decomposition \Eq{eq:auxfield+} is valid because the operator $\hat{o} =  (\hat{c}_i^{\dagger} \hat{c}_j + \hat{c}_j^{\dagger} \hat{c}_i)$ satisfies $ \hat{o}^2  = \hat{o}^4 = \hat{n}_i + \hat{n}_j - 2 \hat{n}_i \hat{n}_j $ when $i \neq j$. Compared to the conventional decompositions routinely employed in the determinantal QMC simulations~\cite{1983PhRvB..28.4059H, 1999PhRvL..82.4155R}, the auxiliary field in~\Eq{eq:auxfield+} couples to  fermion hoppings instead of the density operators~\cite{Scalapino:1984wz, Gubernatis:1985wo}. This is one of the key ingredients to avoiding the sign problem. In retrospect, this choice can be motivated by the Corollary (\ref{eq:corollary}). 


Plugging \Eq{eq:auxfield+} into~\Eq{eq:expansion}, the square bracket becomes a product of exponentials of fermion bilinear operators. The trace therefore acquires an appealing physical meaning: it is the partition function of an imaginary-time dependent noninteracting system, which evolves alternatively under the free part of the original Hamiltonian $\hat{H}_{0}$ and hopping with an  amplitude $\sigma\lambda$ between the sites $i,j$ that belong to different sublattices. Tracing out these free fermions, one obtains  
\begin{widetext}
\begin{equation}
Z = \sum_{k=0}^{\infty} \left(\frac{\Gamma}{2}\right)^{k}\sum_{\braket{i_{1},j_{1}}} \ldots \sum_{\braket{i_{k},j_{k}}} \sum_{\sigma_{1}=\pm} \ldots \sum_{\sigma_{k}=\pm} \int_{0}^{\beta}d \tau_{1} \ldots \int^{\beta}_{\tau_{k-1}} d\tau_{k}  \, \det\left[I +e^{-(\beta-\tau_{k})K}e^{\Lambda^{\sigma_{k}}_{i_{k}j_{k}}}\ldots e^{{\Lambda^{\sigma_{1}}_{i_{1}j_{1}}} } e^{-\tau_{1} K}\right], 
\label{eq:weight}
\end{equation}
\end{widetext}
where the matrix $(\Lambda^{\sigma}_{ij})_{lm} = \sigma \lambda (\delta_{li}\delta_{mj} + \delta_{lj}\delta_{mi})$ according to the exponential factor of \Eq{eq:auxfield+}. 
Equation~(\ref{eq:weight}) is in the general form of \Eq{eq:general} and the matrix determinant has the form of~\Eq{eq:corollary}. The interaction vertex $e^{\Lambda^{\sigma}_{ij}}$ performs a hyperbolic rotation $\left(\begin{array}{cc}\cosh\lambda & \sigma \sinh \lambda \\ \sigma \sinh\lambda & \cosh\lambda \end{array}\right)$ in the relevant $2\times 2$ block involving the sites $i,j$. Importantly, both the original hopping matrix $K$ and the auxiliary Hamiltonian matrix $\Lambda^{\sigma}_{ij}$ satisfy the condition of the Corollary~(\ref{eq:corollary}). The weight (\ref{eq:weight}) is therefore nonnegative and there is no sign problem. The Monte Carlo method can be used to sample the summations over the interaction bonds and the auxiliary fields as well as the integrations over the imaginary times on equal footing, see Refs. \cite{Iazzi:2014vv, Wang:2015tf} for details about efficient Monte Carlo simulation of \Eq{eq:weight}. 



Using the auxiliary field to decouple the interaction vertex is not the only way to formulate a sign-problem-free QMC approach for the model (\ref{eq:tVmodel}). The Theorem~(\ref{eq:theorem--},\ref{eq:theorem+-}) apply to other components of the $O(n,n)$ group and connect the above solution to the solutions based on the continuous-time interaction expansion method (CT-INT)~\cite{Huffman:2014fj, Wang:2014iba} and the related but more efficient LCT-INT method~\cite{Iazzi:2014vv,Wang:2015tf}. These methods correspond to special choices of the shift $\Gamma = -V/4$ \cite{Werner:2010gu} which results in a purely imaginary coupling strength $\lambda = i \pi$ in \Eq{eq:auxfield+}. The vertex matrix $e^{\Lambda^{\sigma}_{ij}}$ thus has the form $\left(\begin{array}{cc} -1 & 0 \\ 0 &-1 \end{array}\right)$ in the relevant $2\times 2$ block independent of the auxiliary field, which is equivalent to rewriting the interaction term $\hat{v}_{ij} = \frac{V}{4}e^{i\pi(\hat{n}_{i}+\hat{n}_{j})}$ in the LCT-INT approach~\cite{Iazzi:2014vv,Wang:2015tf}. The vertex matrix maps the evolution matrix in \Eq{eq:weight} back and forth between the $O^{++}(n,n)$ and the $O^{--}(n,n)$ components, because the sites $i,j$ belong to different sublattices and the vertex matrix flips the signs of both $\det(M_{11})$ and $\det(M_{22})$. The matrix determinant in \Eq{eq:weight} is thus non-positive for an odd number of vertices according to \Eq{eq:theorem--}. However, the negative value $\Gamma=-V/4$ cancels this sign due to a prefactor  $\left(\frac{\Gamma}{2}\right)^{k}$ in the weight. Hence Theorem~(\ref{eq:theorem}) ensures the absence of a sign problem~\footnote{In fact, the QMC simulation is sign-problem-free for any shift $\Gamma \in [-V/4, 0)$, where $\lambda  =i\pi +  \acosh \left(|1 + \frac{V}{2 \Gamma}|\right)$ and the matrix product in \Eq{eq:weight} lives in either $O^{++}(n,n)$ or $O^{--}(n,n)$ component.} in the auxiliary-field-free (L)CT-INT simulations~\cite{Huffman:2014fj, Wang:2014iba, Wang:2015tf}.  

To take full advantage of the Corollary~(\ref{eq:corollary}), one can further consider long-range interactions in the model~\Eq{eq:tVmodel}, e.g., \emph{attractive} interaction between sites belonging to the \emph{same} sublattice~\cite{Huffman:2014fj, Li:2015jf}. We decouple the interactions as
\begin{equation}
 - \hat{v}_{ij} =  \frac{\Gamma}{2} \sum_{\sigma =\pm} \exp\left[\sigma \lambda \left(\hat{c}_i^{\dagger} \hat{c}_j-\hat{c}_j^{\dagger} \hat{c}_i\right)\right]. 
 \label{eq:auxfield-}
\end{equation}
The coupling strength $\lambda = \acos\left(1 + \frac{V}{2 \Gamma}\right)$ is real for \emph{attractive} interactions and any \emph{positive} shift $\Gamma\ge|V|/4$. The effective single-particle Hamiltonian in the exponential of \Eq{eq:auxfield-} is \emph{antisymmetric} and connects sites in the \emph{same} sublattice, thus fulfills the requirement of the Corollary~(\ref{eq:corollary}). There is no sign problem either~\footnote{Moreover, by expanding chemical potential terms on equal footing with the interaction vertices, one can show there is no sign problem in the model \Eq{eq:tVmodel} with a staggered chemical potential~\cite{Huffman:2014fj} using the Theorem (\ref{eq:theorem}). We thank Shailesh Chandrasekharan and Emilie Huffman for pointing out this to us.}. 

Alternatively, in~\Eq{eq:auxfield+} and \Eq{eq:auxfield-} one can split a fermion into two Majorana operators~\cite{Li:2015jf} and identify two complex-conjugate factors in the Monte Carlo weight~\footnote{We found the convention used in \cite{Wei:2014wwb} is more straightforward in achieving this goal.}. It is however clear that the unconventional decoupling in the hopping channels in Eqs.~(\ref{eq:auxfield+},\ref{eq:auxfield-}) to respect the Corollary is the underlying reason of a nonnegative matrix determinant. In light of \Eq{eq:corollary}, rewriting the fermions using Majorana operators is unnecessary in the Monte Carlo simulations. Nevertheless, the Majorana representation~\cite{Li:2015jf} is an ingenious way to prove the Corollary.
 
We have shown that  Theorem~(\ref{eq:theorem}) unifies the recent solutions of the sign problem~\cite{Huffman:2014fj, Li:2015jf} as different choices of the constant shift $\Gamma$. The Corollary~(\ref{eq:corollary}) is particularly instructive as it suggests that one just needs to decompose the original interacting model into free effective Hamiltonians that satisfy the condition of \Eq{eq:corollary} in order to avoid the sign problem. The mechanism of solving the sign problem using \Eq{eq:theorem} and \Eq{eq:corollary} goes beyond the previous understandings based on the TRS principle~\cite{Anonymous:JutiR0Zk, Hands:2000kq, Wu:2005im}. This can be easily seen from the fact that the real eigenvalues of the matrix $I+M$ are not necessarily doubly degenerate as required by the Kramers theorem \footnote{Another way to see it is different from the TRS principle is that the identity matrix in the Theorem~(\ref{eq:theorem}) and Corollary~(\ref{eq:corollary}) are crucial for the (in)equalities to hold, while this is certainly not the case for the considerations based on the time-reversal-symmetry~\cite{Anonymous:JutiR0Zk, Hands:2000kq, Wu:2005im}.}. 


As a further application~\cite{SM} we consider the following \emph{two-flavor} Hubbard model on a bipartite lattice  
\begin{eqnarray}
\hat{H} & = &\sum_{\alpha=\{\uparrow,\downarrow\}} \sum_{ i,j} \hat{c}_{i\alpha}^{\dagger}K^{\alpha}_{ij}\hat{c}_{j\alpha} + \sum_{i} \hat{v}_{i}, \nonumber \\
\hat{v}_{i} & =& U\left( \hat{n}_{i\uparrow}\hat{n}_{i\downarrow} - \frac{\hat{n}_{i\uparrow}+\hat{n}_{i\downarrow}}{2}\right)  -\Gamma, 
\label{eq:Hubbard}
\end{eqnarray}
where the real hopping matrix $K^{\alpha}$ connects the \emph{same} flavor $\alpha$ on \emph{different} sublattices. 
The model~\Eq{eq:Hubbard} covers a variety of interesting physical systems that were previously inaccessible for determinantal QMC simulations. For example, the choice $K^{\downarrow}=r K^{\uparrow}$ with a ratio $0< r < 1$ realizes the asymmetric Hubbard model, which was implemented  recently in a one-dimensional optical lattice with a tunable ratio $r$~\cite{Jotzu:2015tq}. On the other hand, one can also choose to have spatially anisotropic hopping amplitudes for each flavor, therefore to realize Hubbard models with mismatched Fermi surfaces~\cite{Gukelberger:2014db}. 

All these cases break the SU(2) spin symmetry as well as the time-reversal symmetry, therefore are not guaranteed to be sign-problem-free according to the conventional TRS principle~\cite{Anonymous:JutiR0Zk, Hands:2000kq, Wu:2005im}. However, one can now solve the sign problem using the insights provided by the Corollary (\ref{eq:corollary}). We first consider the $U>0$ case for simplicity. Enlightened by the new understanding, we decouple the interaction term  \Eq{eq:Hubbard} similarly to \Eq{eq:auxfield+} and obtain an auxiliary field coupled to the local spin-flip $(\hat{c}^{\dagger}_{i\uparrow} \hat{c}_{i\downarrow} + \hat{c}_{i\downarrow}^{\dagger} \hat{c}_{i\uparrow} )$, 
which connects \emph{different} flavors on the \emph{same} site. Thus for the ordering of the spin-orbital ($\mathcal{A}\uparrow$, $\mathcal{B}\downarrow$; $\mathcal{B}\uparrow$, $\mathcal{A}\downarrow$), it is easy to see the effective Hamiltonians are bipartite and symmetric, therefore satisfy the condition of the Corollary. This shows  that an auxiliary field coupled to the $x$-component of the spin operator is sign-problem free for the model~(\ref{eq:Hubbard})~\footnote{Such a decomposition was previously discussed in~\cite{Chen:1992wo}. It is also related to the anomalous decoupling used in~\cite{Batrouni:1990ug} up to a particle-hole transformation.}. The attractive case can be studied without a sign problem  by performing a particle-hole transformation to the model. Alternatively, one can perform the decomposition according to \Eq{eq:auxfield-} for attractive interactions, thus have a sign-problem-free simulation with the auxiliary field coupled to the $y$-component of the spin operator. Moreover, there is no sign problem even when we explicitly add spin-flip terms in the Hamiltonian as long as the hopping matrix satisfies the condition of~\Eq{eq:corollary}. This covers a large class of compass Hubbard models~\cite{Nussinov:2015dp} which are relevant to multiorbital and ultracold atoms systems~\cite{Zhao:2008ev, Wu:2008ea, Budich:2015vc}. 

Using the special choice of $\Gamma = -U/4$, the above solution reduces to the (L)CT-INT formulation and the determinant of the two flavors factorizes into two parts 
in the absence of the single-particle spin-flip terms.
Even though the two determinants are not necessarily equal due to the broken TRS, Theorem~(\ref{eq:theorem}) ensures that they have the \emph{same sign} because the evolution matrix of the two flavors lie in the same component of $O(n,n)$. In contrast to the case of spinless fermions, the  vertex matrix of $\hat{v}_{i} = \frac{U}{4}e^{i\pi(\hat{n}_{i\uparrow}+\hat{n}_{i\downarrow})}$ can bring the evolution matrix into all four components of the $O(n,n)$ group since each vertex matrix changes the sign of either $\det(M_{11})$ or $\det(M_{22})$ of both flavors. The Monte Carlo weights of odd expansion orders vanish because of \Eq{eq:theorem+-}. Although the matrix size in the LCT-INT simulation is only half of the previously discussed auxiliary field approach, the use of two-vertices insertion/removal updates~\cite{Rubtsov:2005iw} in the Monte Carlo simulation leads to more complicated updates and measurement procedures~\cite{Kozik:2013ji, YHLiu:2015}. The  auxiliary field approach may thus be advantageous. 



These solutions to the sign problem can also be applied to projector QMC methods~\cite{Sugiyama:1986vt, White:1989wh, Wang:2015tf} which sample the ground state wave-function overlap $\braket{\Psi_{T}|e^{-\Theta \hat{H}} |\Psi_{T}}$ instead of the partition function. One can choose the trial-wave function $\ket{\Psi_{T}}$ as the ground state of a single-particle trial Hamiltonian that  fulfills the condition of Corollary (\ref{eq:corollary}) to avoid the sign problem. 

All the sign-problem-free models solved by \Eq{eq:theorem} and \Eq{eq:corollary} in this Letter are at half-filling on  bipartite lattices with particle number conservation~\footnote{Models with explicit pairing terms naturally fit  the Majorana QMC formalism~\cite{Li:2015jf}.}. It will be interesting to see whether one can even go beyond this constraint. Conversely, we emphasize that the requirements of \Eq{eq:theorem} and \Eq{eq:corollary} are by no means the \emph{necessary} conditions for a sign-problem-free QMC simulation. There should be more ``de-sign'' principles of this kind for fermionic Hamiltonians and quantum Monte Carlo methods. Our work suggests it is fruitful to exploit the inherent Lie group and Lie algebra structure in the Monte Carlo weight to search for such ``de-sign'' principles. Incidentally, both the split orthogonal group and the TRS ``de-sign'' principle seem to be related to the ten-fold way classification of random matrices~\footnote{They solve models in the chiral orthogonal (BDI) and symplectic (AII) symmetry classes respectively.}. It would be interesting to generalize them to other symmetry classes~\cite{zirnbauer1996riemannian, PhysRevB.55.1142, heinzner2005symmetry} and draw connections to recent topological classification of gapped free-fermion systems~\cite{PhysRevB.78.195125, kitaev2009periodic, 1367-2630-12-6-065010}. 

Furthermore, findings reported in this paper apply as well to fermions coupled to quantum spins or $\mathbb{Z}_2$ gauge fields. The Theorem (\ref{eq:theorem}) ensures a matrix determinant with a definite sign after integrating out fermions as long as the split orthogonal group structure is respected. This  allows to design new sign-free models relevant to lattice gauge theories~\cite{SM}. 

\paragraph{Acknowledgements} We  thank Ethan Brown, Peter Br\"ocker, Zi-Xiang Li, Simon Trebst and Hong Yao for useful discussions. We also thank all who contributed to the discussions of the MathOverflow question \href{http://mathoverflow.net/questions/204460/how-to-prove-this-determinant-is-positive}{``How to prove this determinant is positive?''} which led to the proof of the Theorem and the Corollary. The work at ETH was supported by ERC Advanced Grant SIMCOFE, the Swiss National Science Foundation, and the National Center of Competence in Research Quantum Science and Technology QSIT. Gergely Harcos was supported by OTKA grants K 101855 and K 104183 and ERC Advanced Grant 321104.

\bibliographystyle{apsrev4-1}
\bibliography{FermionSign}

\appendix

\clearpage

\section{Proof of \Eq{eq:theorem}: A showcase of the open science movement}
The Theorem~\Eq{eq:theorem} was proved by Gergely Harcos and Terence Tao, which was triggered by Lei Wang's conjecture posted on \href{http://mathoverflow.net/questions/204460/how-to-prove-this-determinant-is-positive}{MathOverflow}. 
Below we review the history of the proof. 

On May 1st 2015, Wang posted a weaker version of the Corollary \Eq{eq:corollary} on MathOverflow formulated as the following
\begin{conjecture*}
Given an arbitrary number of real matrices $A_{i}=\left(\begin{array}{cc} 0 & B_{i} \\ B_{i}^{T} &  0\end{array}\right) $, $\det\left(I+ \prod_{i}e^{A_{i}}\right)\ge 0$ holds. 
\end{conjecture*}
Tao soon pointed out the matrix product $\prod_{i}e^{A_{i}}$ lies in the split orthogonal group. Christian Remling and Will Sawin confirmed this observation and clarified the Lie algebra of the $O(n,n)$ group. A ``counterexample'' to the conjecture (a matrix $M\in O(2,2)$ which has $\det(I+M)<0$) was posted but realized later to lie in the $O^{--}(2,2)$ component. These discussions helped to clarify the topological structure of the split orthogonal group. Harcos then formulated the question in the form of \Eq{eq:theorem++} and outlined a strategy of the proof. Tao finalized the proof and summarized a streamlined version in his blog (Proposition 3 in~\cite{Taoblog}). Wang suggested \Eq{eq:theorem--} and \Eq{eq:theorem+-} as further extensions in the comment section of Tao's blog and got affirmative answers from Tao and Harcos. 
There are other people who contributed to the discussion and confirmation of the proof, some are anonymous users of MathOverflow. 
\section{Three-flavor Hubbard model}
Consider the following \emph{three-flavor} Hubbard model~\cite{Huffman:2014fj}
\begin{eqnarray}
\hat{H} &= &\sum_{\alpha=\{1,2,3\}} \sum_{i,j} \hat{c}_{i\alpha}^{\dagger} K^{\alpha}_{ij} \hat{c}_{j\alpha}+\sum_{i} \sum_{\alpha<\beta} \hat{v}^{\alpha\beta}_{i},\nonumber \\
\hat{v}_{i}^{\alpha\beta} &=&  U_{\alpha\beta}\left( \hat{n}_{i\alpha}\hat{n}_{i\beta} - \frac{\hat{n}_{i\alpha}+\hat{n}_{i\beta}}{2}\right)  -\Gamma_{\alpha\beta}.  
\label{eq:3flavor}
\end{eqnarray}
Again, the hopping amplitude $K^{\alpha}$ is real and bipartite, but it does not need to be the same for each flavor. With \emph{attractive} onsite interactions $U_{\alpha\beta}<0$ a decomposition of the interaction term similar to \Eq{eq:auxfield-} gives an auxiliary field that couples to $(\hat{c}^{\dagger}_{i\alpha} \hat{c}_{i\beta} - \hat{c}_{i\beta}^{\dagger} \hat{c}_{i\alpha})$. For the ordering of the spin-orbital $(\mathcal{A}1, \mathcal{A}2, \mathcal{A}3; \mathcal{B}1, \mathcal{B}2, \mathcal{B}3)$ the exponential factor forms a real antisymmetric matrix that fulfills the condition of the Corollary. There is no sign problem in the simulation. 
Alternatively, similar to the discussions in the main texts,  the determinant factorizes for the three flavors with the choice of $\Gamma_{\alpha\beta}=-U_{\alpha\beta}/4$. Their  product is nonnegative and the formalism reduces the sign-problem free (L)CT-INT simulation~\cite{Huffman:2014fj}. 

~\\
\section{$\mathbb{Z}_{2}$ Lattice gauge theory}
Consider $\mathbb{Z}_{2}$ lattice gauge theory~\cite{RevModPhys.51.659} coupled to $N$-flavors of fermions
\begin{eqnarray}
\hat{H} &= &-\sum_{\braket{i,j}} \hat{\sigma}^{x}_{ij} - g\sum_{plaquettes}\hat{\sigma}^{z}_{ij}\hat{\sigma}^{z}_{jk}\hat{\sigma}^{z}_{kl}\hat{\sigma}^{z}_{li} \nonumber \\ && - t \sum_{\braket{i,j}} \sum_{\alpha=1}^{N} \hat{\sigma}^{z}_{ij}\hat{c}_{i\alpha}^{\dagger} \hat{c}_{j\alpha},
\label{eq:lgt}
\end{eqnarray}
where $\hat{\sigma}^{x}$ and $\hat{\sigma}^{z}$ are Pauli matrices defined on links of a $d$-dimensional bipartite lattice. The fermions are at half-filling and live on lattice sites. The first line of \Eq{eq:lgt} describes the $\mathbb{Z}_{2}$ lattice gauge theory in $d+1$ dimensions. The second line couples the $\mathbb{Z}_{2}$ gauge fields to fermion hoppings across the links. The model is sign problem free because after integrating out the fermions the matrix determinants satisfy the Corollary (\ref{eq:corollary}). Notice that the Monte Carlo weight is nonnegative for \emph{any flavor} $N$  because one does not rely on cancellation of signs between different flavors. 

\end{document}